\documentclass[12pt]{article}

\usepackage{aaspp4}
\usepackage{psfig}
\usepackage{floatflt}

\newcommand{\ffffff}[1]{\mbox{$#1$}}
\newcommand{\scnd}{\mbox{\ffffff{''}\hskip-0.3em.}}

\lefthead{S. L. Ellison et al.}
\righthead{HIRES Spectrum of APM~08279+5255}

\begin{document}
\newcommand{\lya}{Ly$\alpha$}
\newcommand{\lyb}{Ly$\beta$}
\newcommand{\lyg}{Ly$\gamma$}
\newcommand{\HI}{H~\textsc{I}}
\newcommand{\CIV}{C~\textsc{IV}}
\newcommand{\SiIV}{Si~\textsc{IV}}
\newcommand{\kms}{km s$^{-1}$} 
\newcommand{\APM}{APM~08279+5255}
\newcommand{\ch}{$N$(C~IV)/$N$(H~I)}

\def\ltsima{$\; \buildrel < \over \sim \;$}
\def\simlt{\lower.5ex\hbox{\ltsima}}
\def\gtsima{$\; \buildrel > \over \sim \;$}
\def\simgt{\lower.5ex\hbox{\gtsima}}

\title{  KECK HIRES  Spectroscopy of  APM~08279+5255\footnote{The data
presented herein were obtained at the W. M. Keck Observatory, which is
operated as a scientific partnership among the California Institute of
Technology, the University of  California and the National Aeronautics
and Space  Administration.  The Observatory  was made possible  by the
generous financial support of the W. M. Keck Foundation.}}

\author{
Sara L. Ellison\altaffilmark{2}, 
Geraint F. Lewis\altaffilmark{3}, 
Max Pettini\altaffilmark{2}, 
Wallace L. W. Sargent\altaffilmark{4},
Frederic H. Chaffee\altaffilmark{5}, 
Craig B. Foltz\altaffilmark{6}, 
Michael Rauch\altaffilmark{7}, 
Mike J. Irwin\altaffilmark{2}}

\altaffiltext{2}{
Institute of Astronomy, Madingley Road, Cambridge CB3 0HA, UK \nl
Electronic mail contact: {\tt sara@ast.cam.ac.uk}}

\altaffiltext{3}{ Fellow of the Pacific Institute of Mathematical
Sciences 1998-1999, \nl Dept. of Physics and Astronomy, University of
Victoria, PO Box 3055, Victoria, B.C., V8W 3P6, Canada \nl \&
Astronomy Dept., University of Washington, Box 351580, Seattle,
WA 98195-1580}

\altaffiltext{4}{Palomar Observatory, Caltech 105--24, Pasadena, CA 91125}

\altaffiltext{5}{ W. M. Keck Observatory, 65--1120 Mamalahoa Hwy,
Kamuela, HI 96743}

\altaffiltext{6}{MMT Observatory,  University of Arizona, Tucson, AZ 85721}

\altaffiltext{7}{European Southern Observatory, Karl-Schwarzschild-Str. 2, 
D-85748 Garching bei Munchen, Germany.}

\begin{abstract}
\noindent  With an optical  $R$-band magnitude  of 15.2,  the recently
discovered  $z$=3.911 BAL  quasar APM~08279+5255  is  an exceptionally
bright high redshift source. Its  brightness has allowed us to acquire
a high signal-to-noise ratio ${\rm (\sim80)}$, high resolution ($\sim$
6  \kms) spectrum  using the  HIRES echelle  spectrograph on  the 10-m
Keck~I telescope.   Given the quality of the  data, these observations
provide an unprecedented view of associated and intervening absorption
systems. Here  we announce  the availability of  this spectrum  to the
general astronomical community and present a brief analysis of some of
its main features.
\end{abstract}

\keywords{quasars: absorption lines -- quasars: individual (APM 08279+5255)}

\section{INTRODUCTION}\label{introduction}

\noindent  Discovered serendipitously  in  a survey  of Galactic  halo
carbon  stars,  the recently  identified  $z$=3.911  BAL quasar  \APM\
(\cite{ilt98})  possesses an inferred  intrinsic luminosity  of $\sim5
\times  10^{15} L_\odot$ ($\Omega_0=1,  h=0.5$), making  it apparently
the most  luminous system currently  known. A significant  fraction of
this prodigious  emission occurs  at infrared and  sub-mm wavelengths,
arising  in a  massive  quantity of  warm  dust (\cite{le98}).  Recent
observations have further probed  this unusual system; CO observations
have  demonstrated  that \APM\  also  possesses  a  large quantity  of
molecular gas  (\cite{dnw99}), a  reservoir for star  formation, while
the  internal structure  of \APM\  has been  probed  with polarization
studies,  indicating  that  several  lines of  sight  through  various
absorbing  and  scattering regions  are  responsible  for the  complex
polarized spectrum (\cite{hi99}).

Observations  with the  1.0 m  Jacobus Kapteyn  telescope on  La Palma
suggest that  \APM\ is not a  simple point-like source,  but is better
represented  by  a  pair   of  sources  separated  by  $\sim  0\scnd4$
(\cite{ilt98}).  This  was confirmed  with  images  acquired with  the
Canada-France-Hawaii AO/Bonnette which  revealed two images, separated
by  $0\scnd35\pm0\scnd02$, with  an intensity  ratio  of $1.21\pm0.25$
(\cite{ltp98});  such a configuration  is indicative  of gravitational
lensing and  suggests that  our view of  \APM\ has  been significantly
enhanced.   More  recent  NICMOS  images  have  further  refined  this
picture, revealing the presence of a third image between the other two
(\cite{ib99}).  The resulting magnification is by a factor of $\sim70$
for  the point-like  quasar source.  However, even  when gravitational
lensing is taken into account, \APM\ is still one of the most luminous
known QSOs.

We have obtained a high S/N ($\sim 80$), high resolution (6 \kms) 
spectrum of \APM, the result of almost 9 hours of observations with
HIRES at the Keck I telescope. In this paper we describe
some of the most important characteristics of the spectrum and
announce its availability to the general astronomical community.
In a separate paper (\cite{sara99}) we have used these data
to throw new light on the question of the C abundance in low
column density \lya\ forest clouds.

The outline of the paper is as follows.  In \S2 we describe the
observations obtained and the data reduction procedure followed
in order to produce the final spectrum.  Section 3 presents a
brief analysis of some of the absorption systems seen in the QSO
spectrum, illustrating the quality and potential of the data.  We
then, in \S4, describe how the data may be obtained from a
permanent, anonymous ftp directory in Cambridge\footnote{Note
that as well as the data files available in Cambridge, a full
ascii spectrum of \APM\ is available in the electronic version of
this PASP Research Note, together with a complete list of \lya\ forest
fit parameters---see \S3 and Tables 2 and 3} before summarising
the main results of the paper in \S5.\\

\section{OBSERVATIONS AND INITIAL REDUCTION}\label{obs}

\noindent The brightness of \APM\ presents an excellent opportunity to
study  spectroscopically the  intervening absorption  systems  and the
Broad Absorption Lines intrinsic to the QSO. To this end, a program of
high resolution observations was  mounted on the 10-m Keck~I telescope
in Hawaii in April and  May 1998 using HIRES, the echelle spectrograph
at the Nasmyth focus (\cite{v92}).  Data were collected for a total of
31\,500  seconds with  the cross  disperser  and echelle  angles in  a
variety  of  settings  so  as  to obtain  almost  complete  wavelength
coverage  from 4400  to 9250  \AA. A  journal of  the  observations is
presented in Table  1.  The data were reduced  with Tom Barlow's HIRES
reduction  package  (Barlow  1999,  in  preparation)  which  extracted
sky-subtracted  object spectra  for each  echelle order.   The spectra
were wavelength calibrated by reference to a Th-Ar hollow cathode lamp
and mapped onto a  linear, vacuum-heliocentric wavelength scale with a
dispersion  of  0.04  \AA\  per  wavelength  bin.   No  absolute  flux
calibration  was  performed,   although  standard  star  spectra  were
obtained and are available (see \S4 below).  Lastly, the orders of the
individual  2-D  spectra and  corresponding  sigma  error arrays  were
merged and then co-added with a weight proportional to their S/N.

The  final spectrum has  a resolution  of 6  \kms\ FWHM,  sampled with
$\sim$  3.5 wavelength bins,  and S/N  between 30  and 150.   The full
spectrum is presented in Table~2,  and in graphical format in Figure~1
(note that while this article and the corresponding PASP
Research Note (paper version) presents only a small portion of
the data, the spectrum in its entirety can be found in the electronic
version of PASP).  Given  the exceptional quality of  the data
and  the  scope  that  they  present  for a  wide  range  of  research
interests,  we  make them  available  to  the astronomical  community.
Details of how to obtain additional material relating to this data are
given in \S4.

\section{INTERVENING ABSORBERS IN THE SPECTRUM OF APM~08279+5255}

\subsection{The \lya\ Forest}

\noindent The rich forest of \lya\ clouds, which is seen as a plethora
of  discrete  absorption lines,  is  caused  by line-of-sight  passage
through structures  such as sheets and filaments  in the intergalactic
medium  (IGM). Hydrodynamical  simulations have  shown that  the \lya\
forest  is a natural  consequence of  the growth  of structure  in the
universe  through hierarchical  clustering  in the  presence  of a  UV
ionizing background  (e.g. \cite{hkw96} ; \cite{bd97}).   For a recent
comprehensive review of the properties  of the \lya\ forest, see Rauch
(1998).

We  fitted Voigt profiles  to the  \lya\ forest  lines using  the line
fitting package  VPFIT (\cite{w87}) which determines  the best fitting
values  of  neutral   hydrogen  column  density  $N$(H~I),  absorption
redshift, $z_{\rm abs}$, and Doppler parameter $b$ ($=\sqrt{2}\sigma$)
for  each absorption  component; the  results are  presented  in Table
3. All   \lya\  lines  within   the  redshift   interval  $3.11<z_{\rm
abs}<3.70$  were  fitted.   The   upper  limit  was  chosen  to  avoid
contamination  of the  sample  by lines  associated  with ejected  QSO
material ($z_{\rm  abs} =  3.70$ corresponds to  the blue edge  of the
broad C~IV absorption trough, at an ejection velocity of $\sim$ 13 100
\kms).  The  lower redshift  limit,  $z_{\rm  abs}  = 3.11$  in  \lya,
corresponds to the onset of  the \lyb\ forest. Within these limits the
line   list   in   Table   3   is  complete   for   column   densities
log~$N$(H~I)$>12.5$.  However,  we consider the values  of $N$(H~I) to
be accurate  only for log~$N$(H~I)$<14.5$  since the fits rely  on the
\lya\ line alone which is saturated beyond this limit (no higher order
Lyman  lines were  included  in  the solution  because  of the  severe
blending of the spectrum below  the wavelength of \lyb\ emission, even
at the high resolution of the HIRES spectra).

The column density distribution in the \lya\ forest can be
represented by a power law of the form
\begin{equation}
n(N)dN = N_0 N^{-\beta}dN
\end{equation}
\noindent  (Rauch 1998  and references  therein).  The  column density
distribution  for the  present sample  is  reproduced in  Figure 2.  A
maximum likelihood fit between $12.5  < {\rm log}~N{\rm (H~I)} < 15.5$
yields a power law index $\beta  = 1.27$ (Figure 3). This is likely to
be a lower limit to true  value of $\beta$ because the line density of
the forest at  these redshifts is sufficiently high  that lines can be
missed  due to  blending. In  other words,  the spectra  are confusion
limited for  weak \lya\  lines. Hu et  al. (1995) used  simulations to
model  this  effect  and  concluded  that incompleteness  sets  in  at
log~$N$(H~I) $\simeq  13.20$ and that  at the lowest  column densities
sampled, log~$N$(H~I)= 12.30  -- 12.60, only one in  four \lya\ clouds
is detected.  If we  adopt the same  incompleteness corrections  as in
Table  3 of  Hu  et al.  (1995), we  deduce  $\beta =  1.39$, in  good
agreement with the value $\beta = 1.46$ reported by these authors over
a similar column density range as that considered here.

An  analysis  of   the  C~IV~$\lambda\lambda  1548,  1550$  absorption
associated with the \lya\ forest has been presented elsewhere (Ellison
et al. 1999).  By fitting profiles to the observed C~IV lines, Ellison
et  al.  (1999)  deduced  a  median  $N$(C~IV)/$N$(H~I)  =  1.4$\times
10^{-3}$  for \lya\  absorbers with  log $N$(H~I)$>$14.5.   Of  the 23
\lya\ clouds within the redshift  interval $3.11 < z_{\rm abs} < 3.70$
which   exhibit   associated   C~IV   absorption,   five   also   show
Si~IV~$\lambda\lambda 1393, 1402$ absorption; an example is reproduced
in Figure  4. Table  4 lists  the parameters of  the profile  fits for
these five absorption systems; the  C~IV and Si~IV systems were fitted
separately (that  is, there was  no attempt to  force a common  fit to
both species).   The values of the  $N$(Si~IV)/$N$(C~IV) ratio deduced
for  the  five  systems  (log~$N$(Si~IV)/$N$(C~IV)  $\simeq  -1.2$  to
$-0.1$)  are typical of  those found  at these  redshifts (Boksenberg,
Sargent, \& Rauch 1998).

\subsection{Mg~II Absorbers}

\noindent  The  data  presented  here   can  be  used  to  search  for
Mg~II~$\lambda\lambda 2796, 2803$ systems at  $z_{\rm abs} > 1$ with a
higher sensitivity than  achieved up to now, formally  to a rest frame
equivalent width detection limit of only  a few m\AA.  On the basis of
the results  by Churchill et al.  (1999) we expect to  find many Mg~II
systems  in our spectrum  and indeed  a first  pass has  revealed nine
systems between $z_{\rm abs} =  1.181$ and 2.066, which are reproduced
in Figure  5. The rest frame equivalent  widths of Mg~II~$\lambda2796$
span the range  from $W_r \simeq 2.5$~\AA\ ($z_{\rm  abs} = 1.181$) to
$W_r = 11$~m\AA\  ($z_{\rm abs} = 1.688$). The  former (see Figure 5a,
top  left-hand panel)  is the  most likely  candidate for  the lensing
galaxy,  given its  strength and  redshift.  On the  other hand,  near
$z_{\rm  abs} =  1.55$  there is  a  complex of  three closely  spaced
absorption  systems, each  in turn  consisting of  multiple components
(Figure 5a,  bottom panel);  with a total  velocity interval  of $\sim
450$~km~s$^{-1}$ such  a configuration may  arise in a  galaxy cluster
which presumably could also contribute to the lensing of the QSO.

Table 5  lists the  absorption line parameters  returned by  VPFIT for
five of the nine Mg~II systems.  We did not attempt to fit the $z_{\rm
abs}   =    1.181$   system    because   the   lines    are   strongly
saturated.  Interestingly, for  the other  three  systems---at $z_{\rm
abs}  = 1.211$,  1.812, and  2.041--- VPFIT  could not  converge  to a
statistically acceptable solution,  in the sense that there  is no set
of  values of  $b$ and  $N$(Mg~II)  which can  reproduce the  observed
profiles  of {\it both}  members of  the doublet.  The problem  can be
appreciated  by considering, for  example, the  $z_{\rm abs}  = 1.211$
system  (Figure 5a, top  right-hand panel).  Here, $\lambda  2796$ and
$\lambda   2803$  have  approximately   the  same   equivalent  width,
indicating that  the lines are saturated  and lie on the  flat part of
the curve of growth, and yet  the residual intensity in the line cores
is $\approx 0.45$\,.

We  believe that  the  reason for  this  apparent puzzle  lies in  the
gravitationally lensed  nature of APM~08279+5255. Our  spectrum is the
superposition  of  two  sight-lines   separated  by  0.35  arcsec  and
contributing in  almost equal proportions to the  total counts (Ledoux
et al. 1998). If there  are significant differences in the strength of
Mg~II  absorption between  the two  sight-lines with---in  the example
considered  here,  saturated  absorption  along  one and  weak  or  no
absorption  along the  other---the  composite spectrum  would  have the
character seen in  our data. 

Assuming the lens to be at $z_{\rm lens} = 1.181$ and an
Einstein-de Sitter universe, the three absorption redshifts
$z_{\rm abs}  = 1.211$, 1.812, and 2.041 correspond to tranverse
distances between the two sight-lines (at an angular separation 
of 0.35 arcseconds) of 1.5, 0.75 and 0.59~$h^{-1}$~kpc
respectively. Some may be surprised to find large changes in the
character of the absorption across such small distances, much
smaller than the scales over which the overall kinematics of
galactic halos vary (e.g. Weisheit \& Collins 1976). In
reality, micro-structure in low-ionization absorption lines
is not unusual and has already been seen (even over sub-parsec
scales) in the interstellar medium of the Milky Way (e.g.
Lauroesch et al. 1998 and references therein), of the LMC
(Spyromilio et al. 1995), and of the absorbing galaxy at $z_{\rm
abs} = 3.538$ in front of another gravitationally lensed QSO,
Q1422+231 (Rauch, Sargent, \& Barlow 1999). These authors have
recently reported {\it spatially resolved}\/ HIRES observations
of images A and C of this bright QSO, which are separated by 1.3
arcseconds.

While in our case it is not possible to deconvolve the individual
contributions of the two sight-lines to our blended spectrum,
because in general there is not a unique `solution' to the
composite Mg~II absorption profiles, the data presented here
provide a strong incentive to observe APM~08279+5255
spectroscopically with STIS on the {\it HST}.  Our prediction is
that  Mg~II absorption at $z_{\rm abs}  = 1.211$, 1.812, and
2.041  will exhibit significant differences  between sight-lines
A and B, and that such differences can be used to  probe in fine
detail the spatial structure of low ionization QSO absorbers,
complementing the results of Rauch et al. (1999) on Q1422+231\,.

\section{Obtaining the data} 

\noindent 
The  data  presented  in  this paper are  also  available  in
electronic form at:
\begin{center} {\tt
ftp://ftp.ast.cam.ac.uk/pub/papers/APM08279} 
\end{center} 
As well as the quasar spectrum, which is presented in fits format with
it  associated error  arrays, this  site also  contains  standard star
spectra (2-D), gzipped postscript  plots of the complete QSO spectrum,
an ascii file of a low  resolution spectrum of \APM\ and a README file
containing  all other  relevant information  required for  using these
data. Any questions regarding the data  can be addressed to SLE in the
first instance.

We ask that any publications  resulting from analyses of this spectrum
fully acknowledge  the W.~M.~Keck Observatory and  Foundation with the
standard pro forma, listed as a footnote on page 1, and reference this
PASP Research Note as the source of the spectrum.

\section{Summary}

\noindent We have presented a brief analysis of the absorption systems
seen in the  HIRES echelle spectrum of the  gravitationally lensed BAL
QSO APM~08279+5255. The \lya\  forest was analysed with Voigt profiles
within  a  region  ($3.11<z_{\rm  abs}<3.70$)  deemed to  be  free  of
contamination  from   higher  order   Lyman  lines  and   ejected  QSO
material.  The H~I  column density  distribution is  well fitted  by a
power law  with slope $\beta =  1.27$ between log~$N$(H~I)  = 12.5 and
15.5; a  higher value, $\beta =  1.39$, is obtained  when allowance is
made  for  line  confusion  at  the  low column  density  end  of  the
distribution.  Approximately   half  of  the  \lya\   lines  with  log
$N$(H~I)$>14.5$ have associated C~IV absorption (Ellison et al. 1999);
five   of   these  C~IV   systems   also   show   Si~IV  with   ratios
$N$(Si~IV)/$N$(C~IV) between $\approx 1$ and $\approx 1/15$.

We identified  nine Mg~II  systems between $z_{\rm  abs} =  1.181$ and
2.066 two of  which are candidates for absorption  associated with the
lens.  For  three  Mg~II  systems  we infer  that  there  are  spatial
differences in the absorption between  the light-paths to the two main
images  of the  QSO (which  are unresolved  in our  study).  Given the
exceptional brightness of  APM~08279+5255, the spectrum presented here
is among  the best ever obtained for  a high redshift QSO;  we make it
available  to the astronomical  community so  that it  can be  used in
conjunction with  other forthcoming studies of  this remarkable object
and sightline.

\newpage

\begin{deluxetable}{cccc}
\footnotesize \tablecaption{
JOURNAL OF HIRES OBSERVATIONS OF APM~08279+5255}
\tablewidth{400pt}
\tablehead{
\colhead{Date} & 
\colhead{Integration time (s)} &  
\colhead{Wavelength range (\AA)} &
\colhead{Typical S/N}
}
\startdata
April 1998 & 1800 & 4400 -- 5945 & 15 \nl
April 1998 & 1800 & 4400 -- 5945 & 15 \nl
April 1998 & 1800 & 4400 -- 5945 & 15 \nl
April 1998 & 1800 & 4400 -- 5945 & 15 \nl
May 1998   & 2700 & 4410 -- 5950 & 25 \nl
May 1998   & 2700 & 4400 -- 5945 & 25 \nl
May 1998   & 900  & 5440 -- 7900 & 30 \nl
May 1998   & 3000 & 5440 -- 7900 & 60 \nl
May 1998   & 3000 & 5440 -- 7900 & 60 \nl
May 1998   & 3000 & 5475 -- 7830 & 55 \nl
May 1998   & 3000 & 5475 -- 7830 & 55 \nl
May 1998   & 3000 & 6765 -- 9150 & 50 \nl
May 1998   & 3000 & 6850 -- 9250 & 50 \nl
& & & \nl
Summed Total & 31\,500 & 4400 -- 9250 & 30 --150 \nl
\enddata
\end{deluxetable}

\newpage

\begin{deluxetable}{ccc}
\footnotesize \tablecaption{
ASCII LISTING OF THE HIRES SPECTRUM OF APM~08279+5255 SHOWN IN
FIGURE 1\,$^{a}$}
\tablewidth{400pt}
\tablehead{
\colhead{Wavelength (\AA)} &  
\colhead{Data (counts)} & 
\colhead{1$\sigma$ error} 
}
\startdata
5700.00 & 337.59 & 7.29 \nl
5700.04 & 319.87 & 7.13 \nl
5700.08 & 296.68 & 6.93 \nl
5700.12 & 281.57 & 6.77 \nl
5700.16 & 265.92 & 6.66 \nl
5700.20 & 246.77 & 6.45 \nl
5700.24 & 233.27 & 6.34 \nl
5700.28 & 238.33 & 6.39 \nl
5700.32 & 217.76 & 6.19 \nl
5700.36 & 219.35 & 6.19 \nl
\enddata
\tablenotetext{a}{The full version of the Table is available
in the electronic version of the PASP Research Note; only the first
10 lines are reproduced in this printed version.}
\end{deluxetable}

\begin{deluxetable}{lll}
\footnotesize \tablecaption{
VOIGT PROFILE PARAMETER FITS FOR \lya\ LINES WITH log~$N$(H~I)$>$12.5
AND $3.11<z_{\rm abs}<3.70$\,$^{a}$}
\tablewidth{400pt}
\tablehead{
\colhead{Redshift, $z_{\rm abs}^{b}$} & 
\colhead{log $N$(H~I)$^c$} &  
\colhead{$b$ value (\kms)$^d$} 
}
\startdata
3.14990 & 13.51$^3$ & 24.8$^1$ \nl
3.15073 & 12.75$^3$ & 12.2 \nl
3.15109 & 12.55$^3$ & 36.5$^3$ \nl
3.15366 & 13.93 & 42.0 \nl
3.15465 & 13.35 & 36.0 \nl
3.15637 & 12.95 & 16.5 \nl
3.15741 & 12.70 & 7.0$^1$ \nl
3.15783 & 12.92$^3$ & 7.6$^1$  \nl
3.15840 & 14.10 & 33.1 \nl
3.15930 & 14.43 & 20.9 \nl
\enddata
\tablenotetext{a}{The full version of the Table is available
in the electronic version of the PASP Research Note; only the first
10 lines are reproduced in this printed version.  The error
designations below also apply to the flags in the full version of this
table.}
\tablenotetext{b}{Redshift error $= \pm 10^{-5}$}
\tablenotetext{c}{$N$(HI) error $<$ 30\% unless otherwise stated}
\tablenotetext{d}{Doppler parameter error $<$ 10\% unless otherwise stated}

\tablenotetext{1}{Error $<$ 20\%}
\tablenotetext{2}{Error $<$ 30\%}
\tablenotetext{3}{Error $>$ 30\%}

\end{deluxetable}

\begin{deluxetable}{cccc}
\footnotesize 
\tablecaption{
VOIGT PROFILE FITS FOR METAL LINE SYSTEMS THAT EXHIBIT BOTH C~IV AND Si~IV
ABSORPTION}
\tablewidth{\textwidth}
\tablehead{
\colhead{Transition, X} &
\colhead{Redshift, $z_{\rm abs}^a$} &  
\colhead{log $N$(X)$^b$} & 
\colhead{$b$ value (\kms)$^c$} 
}
\startdata
C~IV  & 3.37677 & 12.24 & 24.5 \nl
C~IV    & 3.37757 & 12.81 & 6.5  \nl
C~IV    & 3.37770 & 13.28 & 19.7  \nl
C~IV    & 3.37882 & 12.37 & 10.8  \nl
C~IV    & 3.37891 & 13.24 & 22.5  \nl
C~IV    & 3.37969 & 13.56 & 45.4  \nl
C~IV    & 3.37969 & 12.89 & 17.0  \nl
Total C~IV   &         & 13.96$\pm0.03$ &       \nl
\\
Si~IV &   3.37857 &  12.36 &     6.6 \nl
Si~IV &   3.37881 &  12.77 &     9.2 \nl
Si~IV &   3.37907 &  12.44 &    10.9 \nl
Si~IV &   3.37948 &  12.21 &    12.8 \nl
Si~IV &   3.37979 &  12.52 &    37.5 \nl
Si~IV &   3.37982 &  11.95 &     7.7 \nl
Si~IV &   3.38027 &  11.41 &     3.1 \nl
Total Si~IV &           &  13.23$\pm0.02$ &   \nl
\hline
      & & & \nl
C~IV  & 3.50125 &  13.01 & 16.8  \nl
C~IV     & 3.50141 & 12.39 & 9.9  \nl
C~IV     & 3.50204 & 12.24 & 5.3  \nl
C~IV     & 3.50209 & 13.08 & 22.6  \nl
C~IV     & 3.50281 & 12.39 &  32.8  \nl
Total C~IV   &         & 13.46$\pm0.04$ &       \nl
\\
Si~IV &   3.50123 & 12.55   &    14.9 \nl
Si~IV &  3.50131  &  12.14  &    2.7 \nl
Si~IV &  3.50146  & 12.35   &    6.7 \nl
Si~IV & 3.50198   & 12.90   &   10.1 \nl
Si~IV &   3.50219 &  12.29  &    7.6 \nl
Si~IV &   3.50237 &  12.71  &    9.9 \nl
Total Si~IV &           &  13.35$\pm0.02$ &   \nl
\hline
	&		&	\nl
	&		&	\nl
	&		&	\nl
	&		&	\nl
	&		&	\nl
	&		&	\nl
	&		&	\nl
	&		&	\nl
	&		&	\nl
	&		&	\nl
	&		&	\nl
C~IV  & 3.51380 & 12.88 & 17.1  \nl
C~IV   & 3.51436 & 12.54 & 14.8 \nl
Total C~IV   &         & 13.04 $\pm0.03$ &       \nl
\\
Si~IV & 3.51341   & 11.71   &    1.4 \nl
Si~IV &  3.51357  & 11.65   &    3.3 \nl
Si~IV &  3.51376  & 12.58   &    9.3 \nl
Si~IV &  3.51401  & 12.02   &    4.6 \nl
Si~IV &  3.51423  &  11.62  &    5.6 \nl
Total Si~IV &           &  12.79$\pm0.03$ &   \nl
\hline
	&		&	\nl
C~IV  & 3.55811 & 12.85 & 11.5  \nl
C~IV     & 3.55842 & 12.63 & 12.7 \nl
Total C~IV   &         & 13.06$\pm0.03$ &       \nl
\\
Si~IV &  3.55826 &  11.82   &   14.4 \nl
Total Si~IV &           &  11.82$\pm0.10$ &   \nl
\hline
	&		&	\nl
C~IV  & 3.66863 & 12.88 & 89.3 \nl
C~IV   & 3.66892 & 13.22 & 36.9 \nl
C~IV   & 3.67082 & 12.97 & 21.4  \nl
C~IV   & 3.67131 & 12.81 & 15.3 \nl
Total C~IV   &         & 13.60 $\pm0.08$ &       \nl
\\
Si~IV & 3.67022 &  12.22    &    8.0 \nl
Si~IV &   3.67076&  12.36   &    9.8 \nl
Total Si~IV   &         & 12.60$\pm0.05$ &       \nl
\enddata
\vspace{2cm}
\tablenotetext{a}{Redshift error $= \pm 10^{-5}$}
\tablenotetext{b}{Column density error $<$ 40\% for individual
components.  Total error for each metal line system indicated
individually}
\tablenotetext{c}{Doppler parameter error $<$ 15\%}
\end{deluxetable}

\begin{deluxetable}{ccc}
\footnotesize
 \tablecaption{VOIGT PROFILE PARAMETER FITS TO Mg~II SYSTEMS}
\tablewidth{400pt}
\tablehead{
\colhead{Redshift, $z_{abs}^a$} &  
\colhead{log $N$(Mg~II)$^b$} & 
\colhead{$b$ value (\kms)$^c$} 
}
\startdata
 1.29083  & 11.98  &     3.8 \nl   
 1.29101  & 11.58  &    11.6  \nl
Total Mg II & 12.13 $\pm0.05$& \nl
\hline
  &   &  \nl
1.44442  &  11.73  &  3.2 \nl
1.44433  & 11.87  &     13.9 \nl
Total Mg II & 12.11$\pm0.12$ & \nl
\hline
  &  &  \nl
  1.54859 & 11.80  &       8.6 \nl
Total Mg II & 11.80$\pm0.15$ & \nl
& & \nl
  1.54949 & 11.55 &      2.2 \nl
  1.54996 & 12.31 &      4.9 \nl
 1.54959 & 12.04 &       4.3 \nl
 1.54973 & 12.44 &      10.5 \nl
 1.55000 & 11.86 &      15.0 \nl
Total Mg II & 12.84 $\pm0.03$ & \nl
 & & \nl
 1.55226 & 12.26  &       5.2 \nl 
 1.55238 & 12.21  &       4.2 \nl
Total Mg II & 12.54 $\pm0.02$ & \nl
\hline
  &  &  \nl
1.68727  & 11.47  &8.3 \nl
Total Mg II & 11.47 $\pm0.20$ \nl
\hline
  &  &  \nl
 2.06685  & 12.40  &     29.4 \nl
2.06784  & 11.65  &      21.3 \nl
 2.06669  & 11.75  &      4.6 \nl
Total Mg II & 12.55$\pm0.04$ & \nl
\enddata
\tablenotetext{a}{Redshift error $= \pm 10^{-5}$}
\tablenotetext{b}{Column density error $<$ 40\% for individual
components.  Total error for each metal line system indicated
individually}
\tablenotetext{c}{Doppler parameter error $<$ 15\%}
\end{deluxetable}

\vspace{5cm}

\newpage
\begin{figure}[H]
\centerline{
\psfig{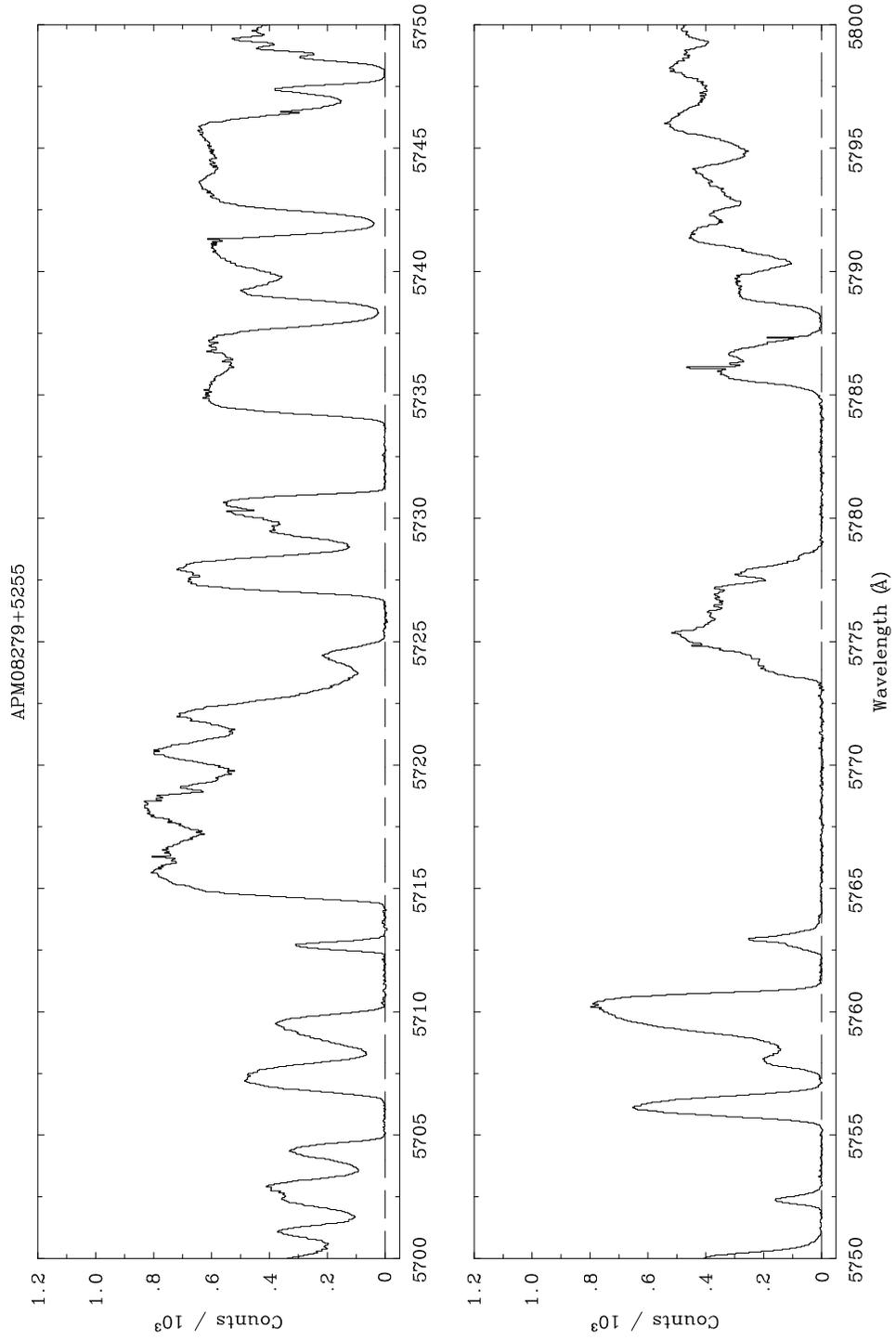}
}
\caption[]{  
A 100 \AA\ section of the HIRES spectrum of \APM, between 5700 and 5800\AA.}
\label{spectrum}
\end{figure}

\begin{figure}[H]
\centerline{
\psfig{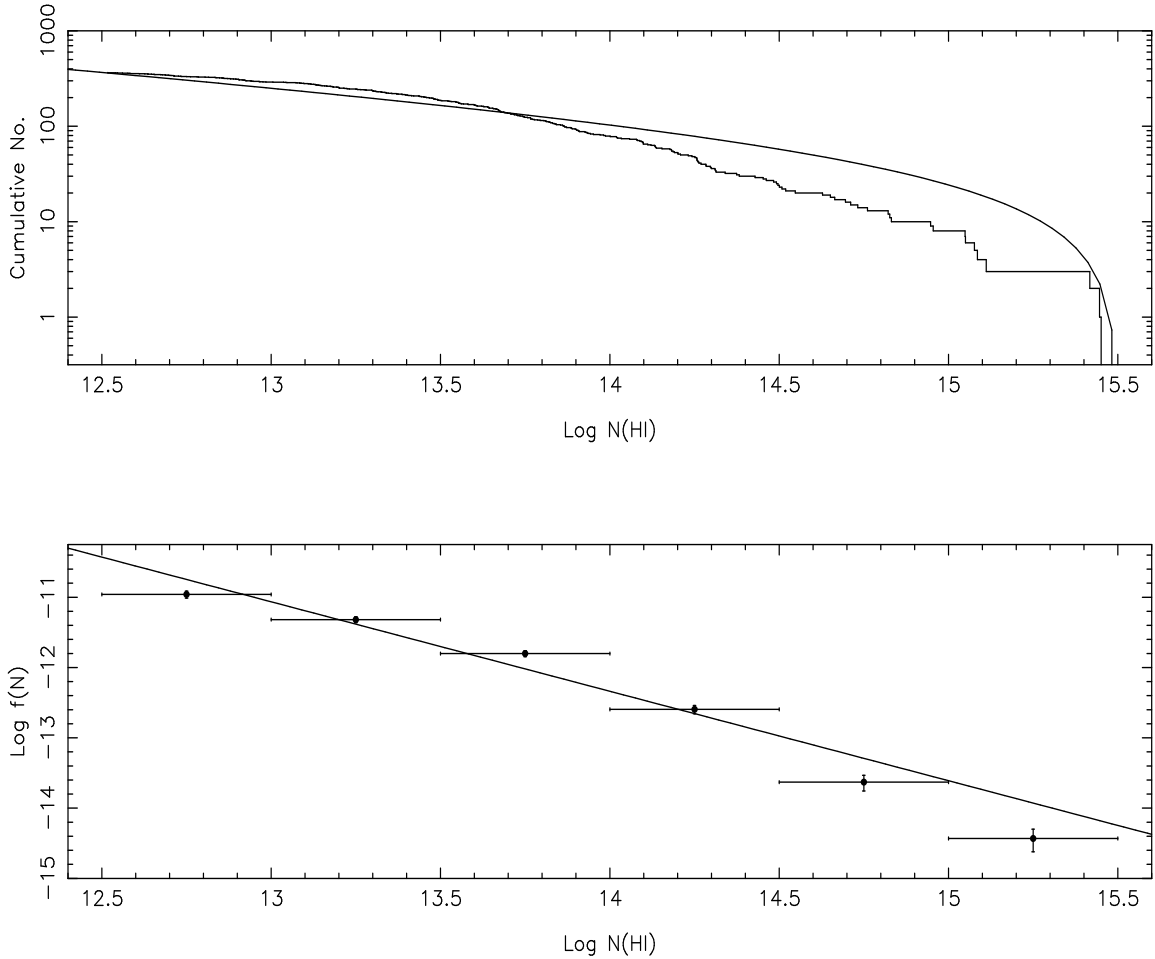}
}
\caption[]{The column density distribution function for the
\lya\ forest in APM~08279+5255, in cumulative (top) and
differential (bottom) forms. The data have been binned for
display purposes only. In each panel the continuous line 
refers to a power law distribution with exponent $\beta = 1.27$
(see Figure 3).}
\end{figure}

\begin{figure}[H]
\centerline{
\psfig{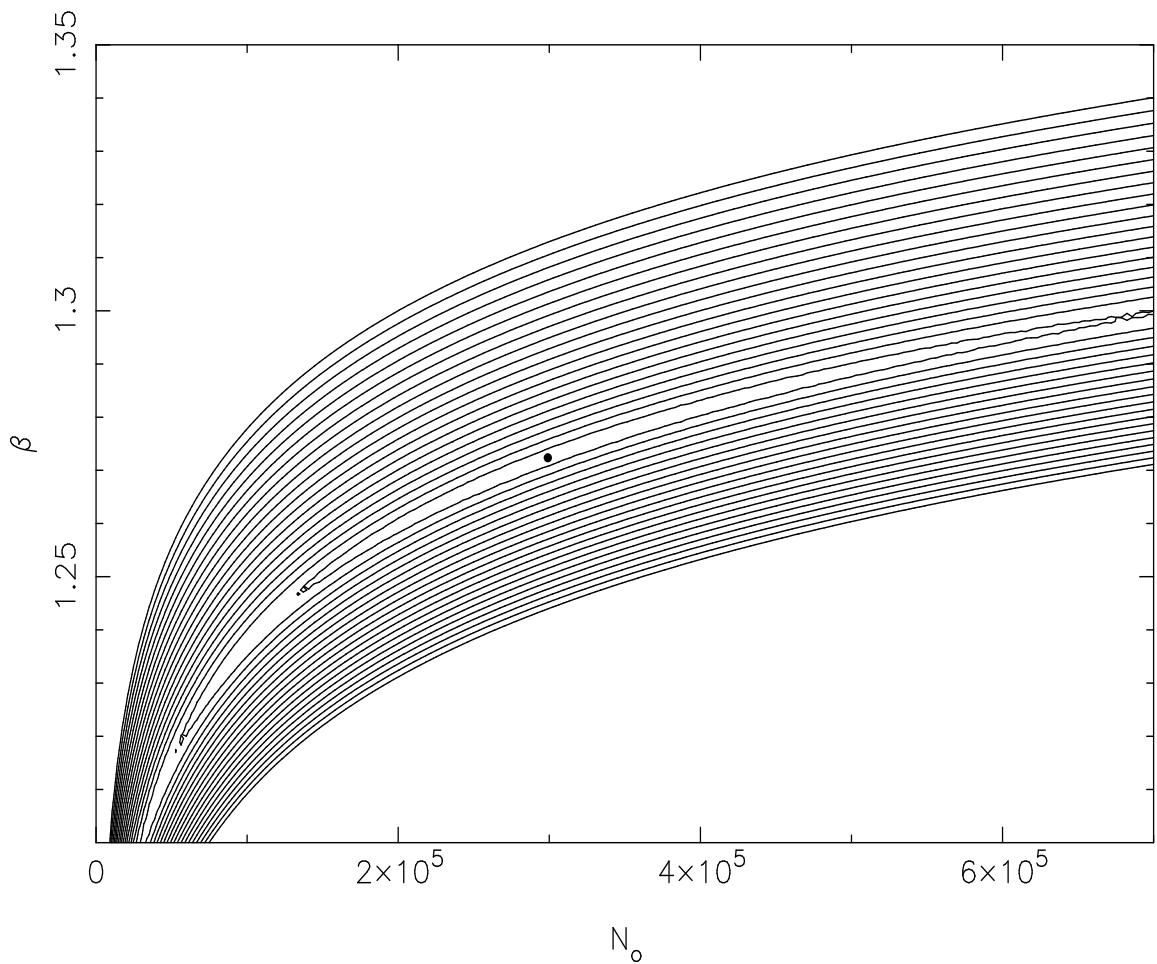}
}
\caption[]{Maximum likelihood contours for the fit to the
$N$(H~I) column density distribution assuming a power law
of the form given in eq. (1).}
\end{figure}

\begin{figure}[H]
\centerline{
\psfig{figure=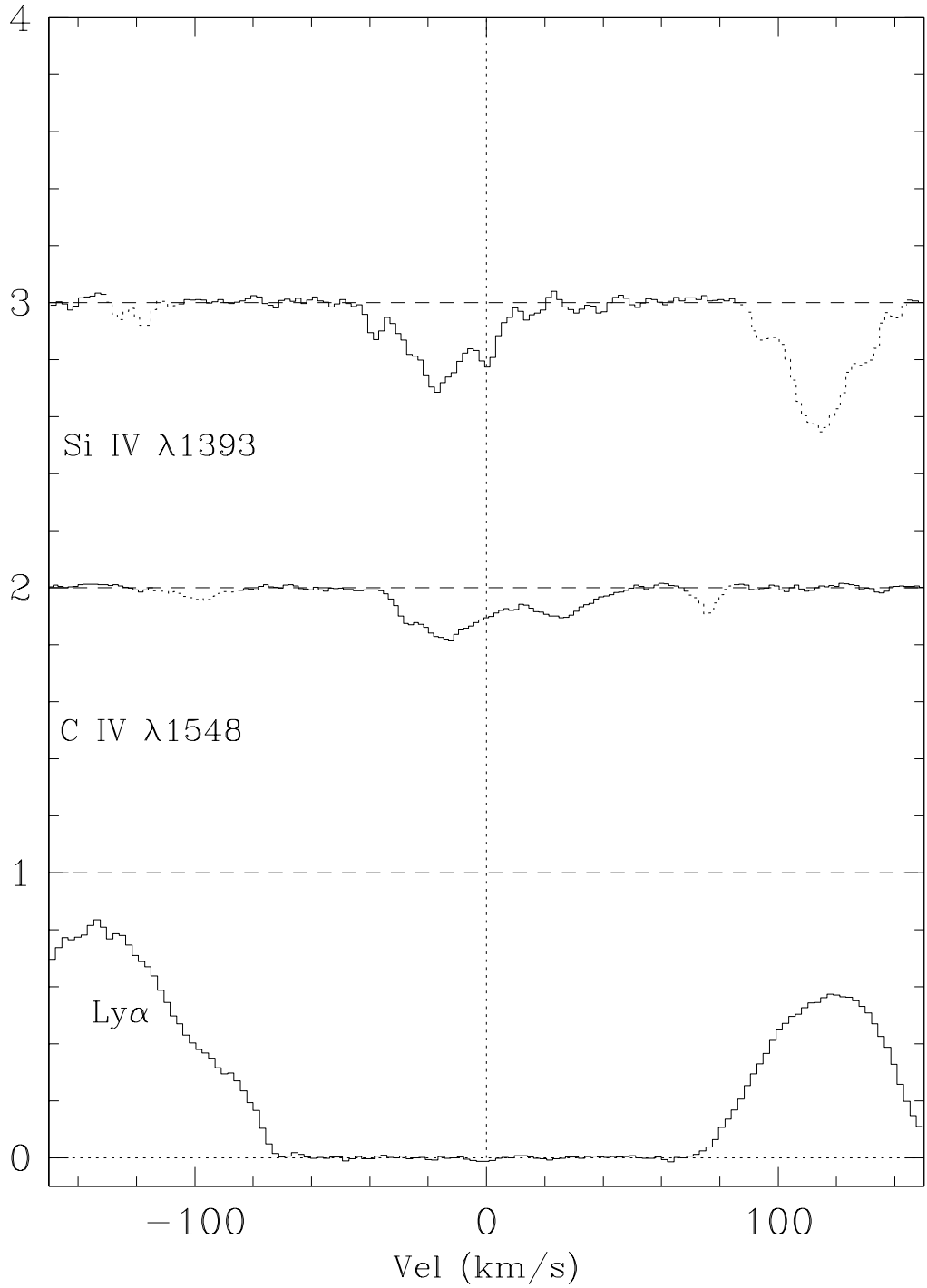,width=6in,angle=0}
}
\caption[]{  
An example of a saturated \lya\ line with associated C~IV and
Si~IV absorption (the scale of the y-axis has been offset for
clarity).  This is the absorption system at $z_{\rm abs} = 3.514$
in Table 4, with log~$N$(C~IV) = 13.04 and log~$N$(Si~IV) =
12.79 (absorption features from other systems are shown with broken
lines).}
\end{figure}

\begin{figure}[H]
\centerline{
\psfig{figure=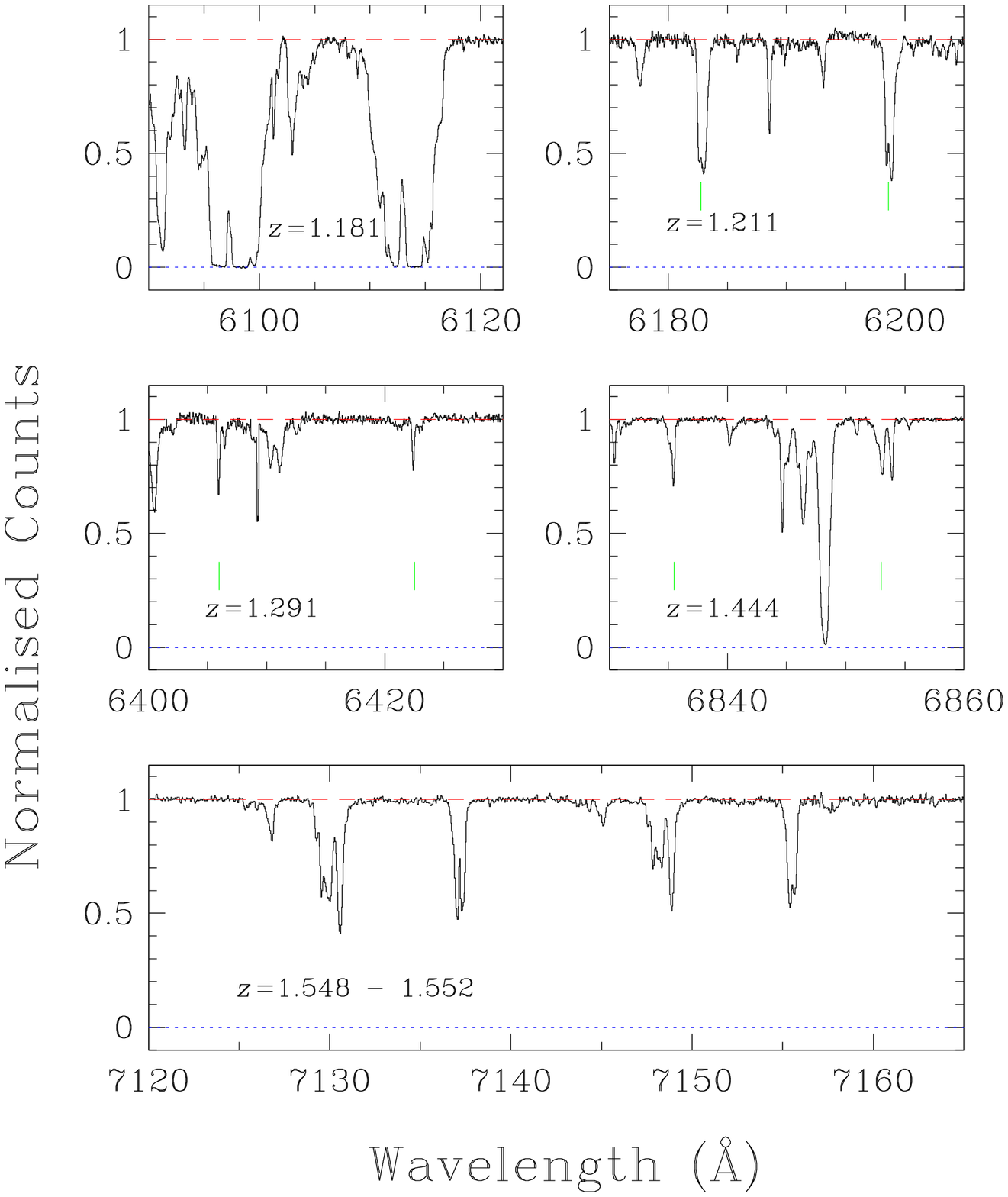,width=7in,angle=0}
}
\caption[]{{\it (a):}~~Mg~II absorbers in the spectrum of APM~08279+5255}
\end{figure}

\addtocounter{figure}{-1}
\begin{figure}[H]
\centerline{
\psfig{figure=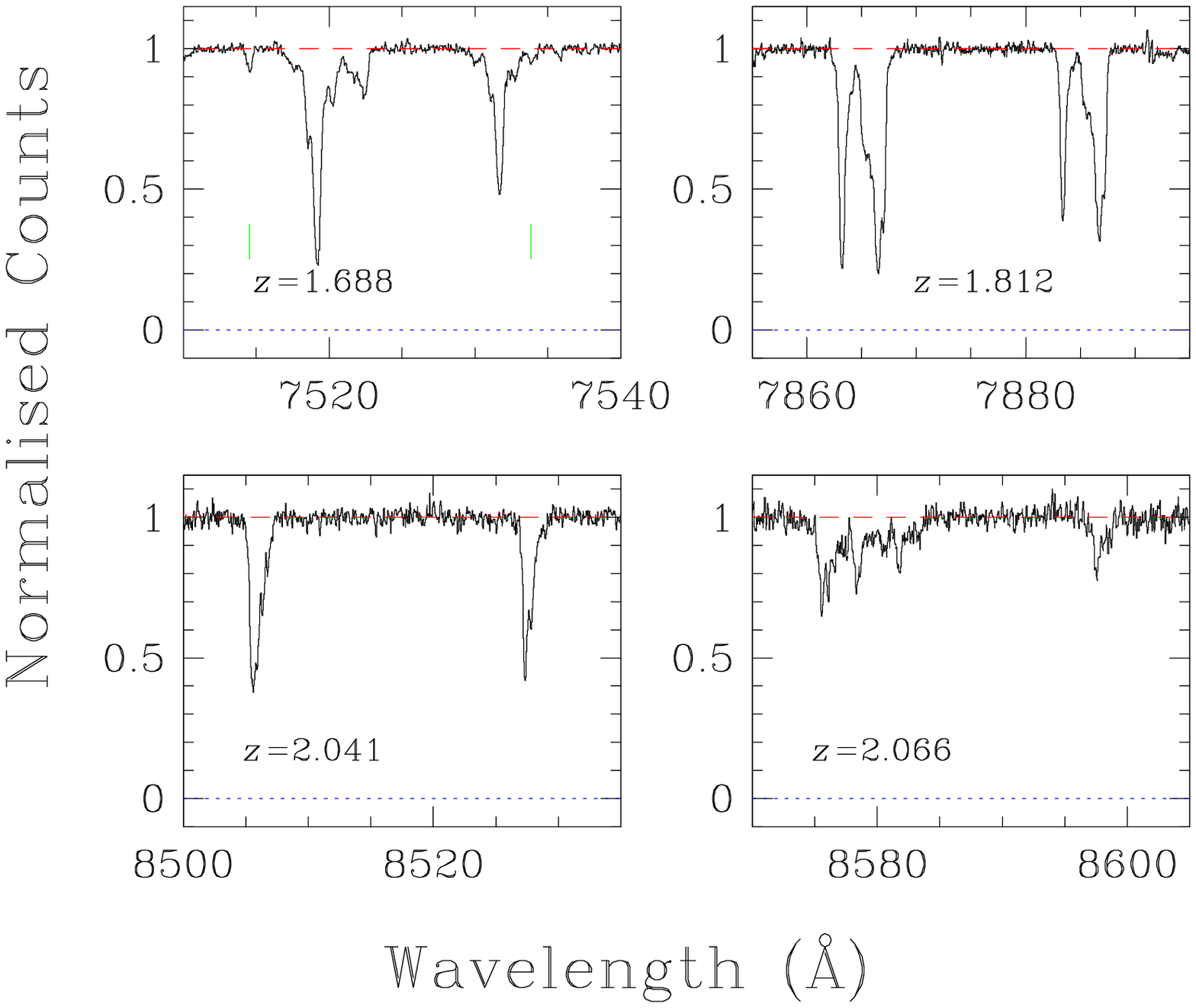,width=7in,angle=0}
}
\caption[]{{\it (b):}~~Mg~II absorbers in the spectrum of
APM~08279+5255}
\end{figure}

\end{document}